\documentstyle[graphicx,aaspp4]{article} 
\begin{document} 
\pagenumbering{arabic}

\newcommand{\kdnode}{n}
\newcommand{\myindex}[1]{\mbox{\scriptsize #1}}
\newcommand{\thingfield}[2]{{#1}.{\myindex{\sc #2}}}
\newcommand{\thingsplitdim}[1]{\thingfield{#1}{SplitDim}}
\newcommand{\thingsplitval}[1]{\thingfield{#1}{SplitValue}}
\newcommand{\mbw}{\varname{MinBoxWidth}}
\newcommand{\varname}[1]{\mbox{\sc #1}}
\newcommand{\thinghr}[1]{\thingfield{#1}{BoundBox}}
\newcommand{\query}{\boldthing{q}}
\newcommand{\boldthing}[1]{\mbox{\bf #1}}
\newcommand{\hisep}{\thingsep{hi}}
\newcommand{\thingsep}[1]{\limitsub{s}{#1}}
\newcommand{\limitsub}[2]{{\mbox{\it #1}}_{#2}}
\newcommand{\kdnodenumpoints}{\thingnumpoints{\kdnode}}
\newcommand{\thingnumpoints}[1]{\thingfield{#1}{NumPoints}}
\newcommand{\rangecount}{\algname{RangeCount}}
\newcommand{\algname}[1]{\mbox{\bf #1}}
\newcommand{\maxdist}{\varname{MaxDist}}
\newcommand{\mindist}{\varname{MinDist}}

\newcommand{\kdtree}{{\it k}d-tree}
\newcommand{\kdtrees}{{\it k}d-trees}
\newcommand{\kdnodesplitdim}{\thingsplitdim{\kdnode}}
\newcommand{\kdnodesplitval}{\thingsplitval{\kdnode}}
\newcommand{\rmin}{r_{\myindex{min}}}
\newcommand{\kdnodehr}{\thinghr{\kdnode}}
\newcommand{\cleft}{\countsub{left}}
\newcommand{\cright}{\countsub{right}}
\newcommand{\countsub}[1]{{\countvar}_{\myindex{#1}}}
\newcommand{\countvar}{C}
\newcommand{\kdnodeleft}{\thingleft{\kdnode}}
\newcommand{\thingleft}[1]{\thingfield{#1}{Left}}
\newcommand{\kdnoderight}{\thingright{\kdnode}}
\newcommand{\thingright}[1]{\thingfield{#1}{Right}}

\title{Computational AstroStatistics: Fast and Efficient Tools for Analysing Huge Astronomical Data Sources}
\author{R. C. Nichol$\footnote{Carnegie Mellon University, 5000 Forbes
Ave, Pittsburgh, PA-15232}$, S. Chong$^1$, A. J. Connolly$\footnote{University
of Pittsburgh, Pittsburgh, PA-15260}$, S. Davies$^1$, C. Genovese$^1$,
A. M. Hopkins$^2$, C. J. Miller$^1$, A. W. Moore$^1$, D. Pelleg$^1$,
G. T. Richards$\footnote{The Pennsylvania State University, University Park,
PA-16802}$, J. Schneider$^1$, I. Szapudi$\footnote{IfA, University of Hawaii,
Honolulu, Hawaii 96822}$, L. Wasserman$^1$
}

\begin{abstract}
I present here a review of past and present multi--disciplinary research of
the Pittsburgh Computational AstroStatistics$\footnote{See
http://www.picagroup.org for a full list of PiCA members and our latest
papers, research and software}$ (PiCA) group. This group is dedicated to
developing fast and efficient statistical algorithms for analysing huge
astronomical data sources. I begin with a short review of multi--resolutional
{\kdtrees} which are the building blocks for many of our algorithms. For
example, quick range queries and fast $N$--point correlation functions. I will
present new results from the use of Mixture Models (Connolly et al. 2000) in
density estimation of multi--color data from the Sloan Digital Sky Survey
(SDSS). Specifically, the selection of quasars and the automated
identification of X--ray sources. I will also present a brief overview of the
False Discovery Rate (FDR) procedure (Miller et al. 2001a) and show how it has
been used in the detection of ``Baryon Wiggles'' in the local galaxy power
spectrum and source identification in radio data. Finally, I will look forward
to new research on an automated Bayes Network anomaly detector and the
possible use of the Locally Linear Embedding algorithm (LLE; Roweis \& Saul
2000) for spectral classification of SDSS spectra.

\end{abstract}

\section{Introduction}
\label{intro}

In this paper, I present an update on the past and present work of the
Pittsburgh Computational AstroStatistics (PiCA) group; a multi--disciplinary
group of researchers from Computer Science, Statistics, and Astrophysics
dedicated to developing fast and efficient algorithms for the analysis of huge
astronomical datasets (see Nichol et al. 2000 a previous review of our
work). The work presented by Larry Wasserman in this volume is part of the
PiCA group research but is not discussed herein for obvious reasons.

The motivation for this work is two--fold. First, the quantity of data being
collected is increasing rapidly and we stand on the threshold of the so-called
``data flood''. By the end of this decade, we will have collected petabytes of
astronomical data {\it e.g.} LSST \& Planck.  The sheer size and
dimensionality of these datasets will restrict our ability to navigate and
analyse these huge data sources and we will need new techniques to help
us. The proposed ``Virtual Observatory'' (VO; see papers by Alex Szalay and
George Djorgovski in this volume) is designed to address the issues of
management, distribution and manipulation of such huge, multi--dimensional
astronomical datasets. In this paper, we focus on the need for new analysis
algorithms since an $N^2$ or $N^3$ algorithm -- where $N$ is the number of
data points -- will no work any longer.

Second, we are entering the realm of high precision astrophysics where the
need to make measurements with higher and higher accuracy will increase (see
recent review by Turner 2001). In cosmology, for example, the next decade
will see the drive to measure the cosmological parameters to an accuracy of a
few percent as well as confidently map the distribution of mass in both the
local and distant universe. The drive for higher precision will greatly
benefit from new statistical tools like those discussed herein and by others
in this volume. In general, these new statistical techniques are
computationally intense -- {\it e.g.} the non-parametric techniques discussed
by Larry Wasserman (this volume) -- and therefore, to gain their potential, we
will need to develop fast and efficient implementations of such algorithms. In
this paper, I present some examples of such implementations.

In Section \ref{trees}, I present a brief review of multi--resolutional
KD--trees which are at the heart of much of our technology. In Section
\ref{tree-apps}, I provide some examples of how such trees can speed--up
simple counting queries. In Section \ref{EM}, I will review Mixture Models and
their use in Astrophysics. In Section \ref{FDR}, I will quickly present a new
statistical tool called False Discovery Rate (FDR) and show two recent
applications of this technique. In Section \ref{Bayes}, I will outline our new
work on a Bayes Network anomaly detector, while in Section \ref{high}, I
present initial results from our research of algorithms for mapping high
dimensional spaces.

\section{Multi--Resolutional KD--trees}
\label{trees}

A multi--resolutional KD--tree ({\kdtree}) is a way of organizing a set of
data-points in $k$-dimensional space in such a way that once built, whenever a
query arrives requesting a list of all points in a neighborhood, the query can
be answered quickly without needing to scan every single point.

The root node of the {\kdtree} owns all the data points.  Each non-leaf-node
has two children, defined by a splitting dimension $\kdnodesplitdim$ and a
splitting value $\kdnodesplitval$. The two children divide their parent's data
points between them, with the left child owning those data points that are
strictly less than the splitting value in the splitting dimension, and the
right child owning the remainder of the parent's data points. 

{\kdtrees} are usually constructed top-down, beginning with the full set of
points and then splitting in the center of the widest dimension. It has been
shown that this splitting criteria -- instead of, say, splitting at the median
of the widest dimension -- produces a more balanced tree which is thus closer
to obtaining the desired $O(log\, N)$ performance (see Moore 1991).  This
produces two child nodes, each with a distinct set of points. This
is then repeated recursively on each of the two child nodes.

A node is declared to be a leaf, and is left unsplit, if the widest dimension
of its bounding box is $\leq$ some threshold, $\mbw$. A node is also left
unsplit if it denotes fewer than some threshold number of points, $\rmin$. A
leaf node has no children, but instead contains a list of $k$-dimensional
vectors: the actual data-points contained in that leaf.  The values $\mbw = 0$
and $\rmin = 1$ would cause the largest {\kdtree} structure because all leaf
nodes would denote single data points. In practice, we set $\mbw$ to $1\%$ of
the range of the data point components and $\rmin$ to around 10. The tree size
and construction thus cost considerably less than these bounds because in
dense regions, tiny leaf nodes are able to summarize dozens of data
points. The operations needed in tree-building are computationally trivial and
therefore, the overhead in constructing the tree is negligible.  Also, once a
tree is built it can be re-used for many different analysis operations.

\section{Example Uses of \kdtrees}
\label{tree-apps}

\subsection{Range Counting and Cached Sufficient Statistics}

One of the most common queries made in Astronomy is: how many objects are
within 1 arcminute (or distance $r$) of a given position. As discussed below,
such a query can be performed very quickly using a {\kdtree}.

The key to the speed of such a query is the decorations of the {\kdtree} with
extra information which we refer to as {\em cached sufficient statistics} (see
Moore \& Lee 1998). Specifically, we can store for each node the bounding box
of all the points it contains (call this box $\kdnodehr$). The implication of
this is that every node must contain two new $k$ dimensional vectors to
represent the lower and upper limits of each dimension of the bounding
box. The range search operation takes two inputs. The first is a
$k$-dimensional vector $\query$ called the {\em query point}. The second is a
separation distance $\hisep$. The operation returns the complete set of points
in the {\kdtree} that lie within distance $\hisep$ of $\query$.  Also, we can
store $\kdnodenumpoints$, which is the number of points contained in each
node. Furthermore, we also store the centroid of all points in a node and
their covariance matrix.

Once we have $\kdnodenumpoints$ and $\kdnodehr$, it is trivial to write an
operation that exactly counts the number of data-points within some range
without explicitly visiting all the data-points.                       .

\begin{itemize}
\item
$\rangecount(\kdnode,\query,\hisep)$\\ 
Returns an integer: the number of
points that are both inside the $\kdnode$ and also within distance $\hisep$ of
$\query$.
\item
Let $\mindist$ := the closest distance from $\query$ to $\kdnodehr$.
\item
If $\mindist \geq \hisep$ then it is impossible that any point in
$\kdnode$ can be within range of the query. So simply return 0.
\item
Let $\maxdist$ := the furthest distance from $\query$ to $\kdnodehr$.
\item
If $\maxdist \leq \hisep$ then every point in $\kdnode$ must be within
range of the query. So simply return $\kdnodenumpoints$.
\item
Else, if $\kdnode$ is a leaf node, we must iterate through all the
data-points in its leaf list. Start a counter at zero. For each point,
find if it is within distance $\hisep$ of $\query$. If so, increment
the counter by one. Return the count once the full list has been
scanned.
\item
Else, $\kdnode$ is not a leaf node. Then:
\begin{itemize}
\item
Let $\cleft := \rangecount(\kdnodeleft,query,\hisep)$
\item
Let $\cright := \rangecount(\kdnoderight,query,\hisep)$
\item
Return $\cleft + \cright$.
\end{itemize}
\end{itemize}

\subsection{Fast $N$--point Correlation Functions}

$N$--point correlation functions have a rich history in Astrophysics and have
been extensively used to characterize the large--scale distribution of matter
in the Universe. Moreover, higher--order correlation
functions will become critically important in this new era of high precision
cosmology as they are important tests of biasing and gaussianity (see Szapudi
et al. 2001).

$N$--point correlation functions are however, computationally intensive to
compute especially for large databases and high values of $N$. We have used a
dual {\kdtree} approach to help solve this problem and provide substantial
speed--ups for calculating the $N$--point correlation functions (see Moore et
al. 2000 \& 2001). We note here that substantial speed--ups can also be
achieved by binning the data into cells and performing the calculation
directly on that grid. This is fine for separations larger than the grid size
but fails as one approaches the resolution of the bin size. Our method is
equivalent to an ``all--pairs'' calculation {\it i.e.} if one had visited all
possible pairs of points in the dataset and binned them appropriately.

For more details on our $N$--point correlation function code, the reader is
referred to Moore et al. (2000 \& 2001) as well as our website {\tt
http://www.autonlab.org/}. We note here that the tree structures discussed
herein are optimal for relatively low dimensional spaces ({\it e.g.} a few
tens of dimensions) and other tree structures like Ball--trees and AD--trees
are better for higher--dimensional spaces (see Moore \& Lee 1998)

\section{Using Mixture Models in Astrophysics}
\label{EM}
In Connolly et al. (2000), we presented the use of Mixture Models of Gaussians
to model the probability density function of multi--dimensional astronomical
data. The reader is referred to Connolly et al. (2000) for a detailed review
of Mixture Models including our fast implementation of the algorithm based
upon the {\kdtree} technology discussed above. In this section, we provide two
recent applications of this technology to the Sloan Digital Sky Survey (SDSS).

\subsection{Finding X--ray Sources}
Even after years of hard work, the number of detected X--ray sources with an
optical identification remains small. For example, the
WGACAT$\footnote{http://wgacat.gsfc.nasa.gov}$, SHARC (Romer et al. 2000) \&
RASS (Voges et al. 1999) X--ray catalogs, which contain hundreds of thousands
of X--rays sources, are still mostly unidentified. This is due to the
laborious nature of the optical follow--up.

This will hopefully change soon primarily due to new optical surveys of the
sky and the approaching VO era which will provide new, automated tools to
assist the user. As a pilot study, we are using the SDSS data and the mixture
model algorithm to help automate the optical identification of X--ray sources.

This is achieved as follows. We first obtain photometric multi--color data
($u'$,$g'$,$r'$,$i'$,$z'$) within 15 arcseconds of 7300 WGACAT and SHARC
sources within the boundaries of the SDSS EDR data (see Stoughton et
al. 2001). This results in 377 matches between an SDSS and X--ray
source. Using these data, we cluster the sources in 4D color--space and thus
determine the probability density function for these sources (the best fit
mixture model contains 33 gaussians). This pdf is then used to determine the
likelihood of any new source being an X--ray source. We plan to extend this
work to include further optical and X--ray information {\it e.g.} the optical
morphology and the ratio of the optical and X--ray fluxes (see Stocke et
al. 1991).  This will facilitate a robust and automatic identification for a
large number of X--ray sources presently lying undiscovered in catalogs like
WGACAT. A preliminary version of this system is available at {\tt
http://ranger.phys.cmu.edu/users/xray/}.

\begin{center}
\includegraphics[width=2.3in]{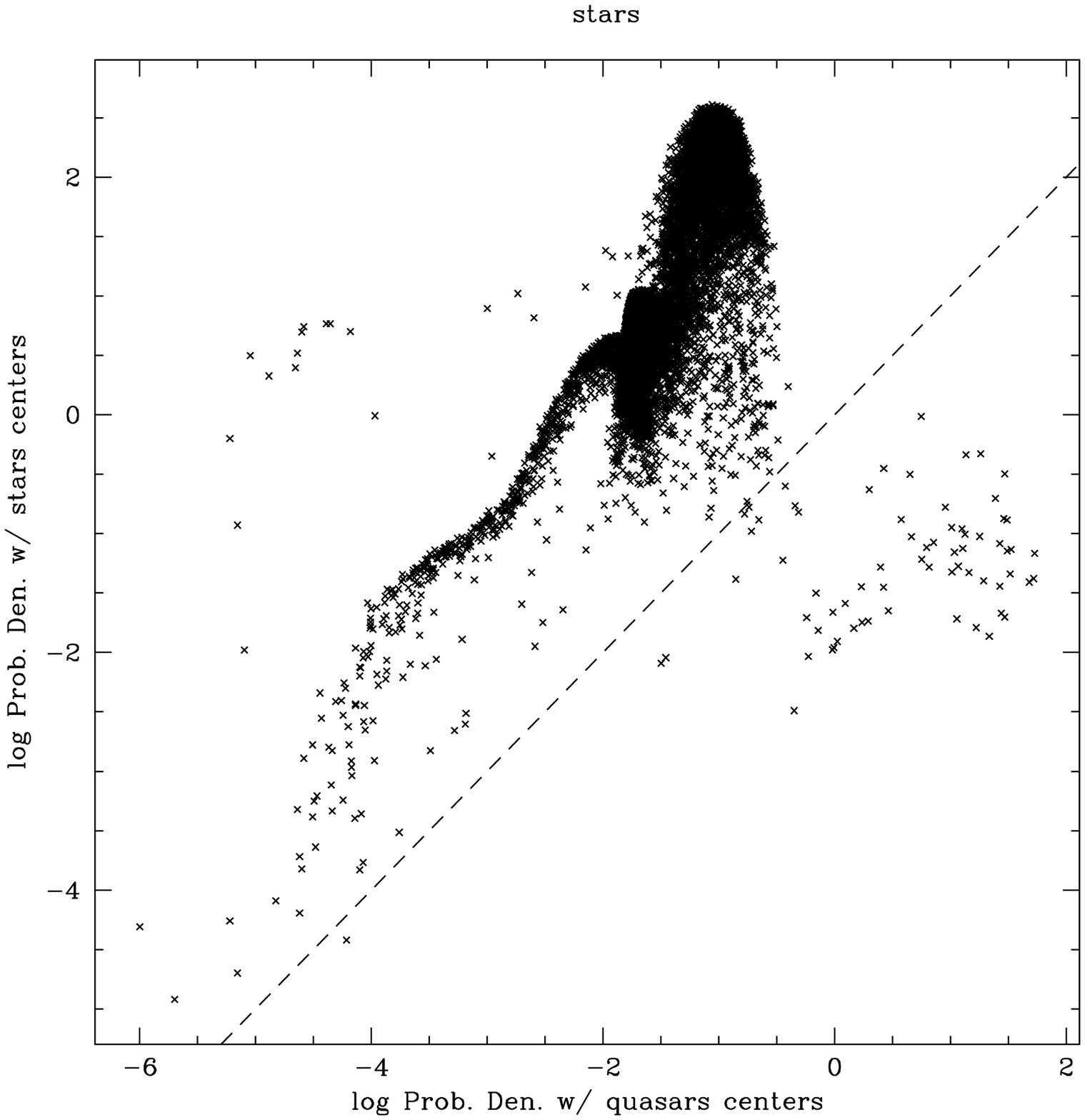}\includegraphics[width=2.3in]{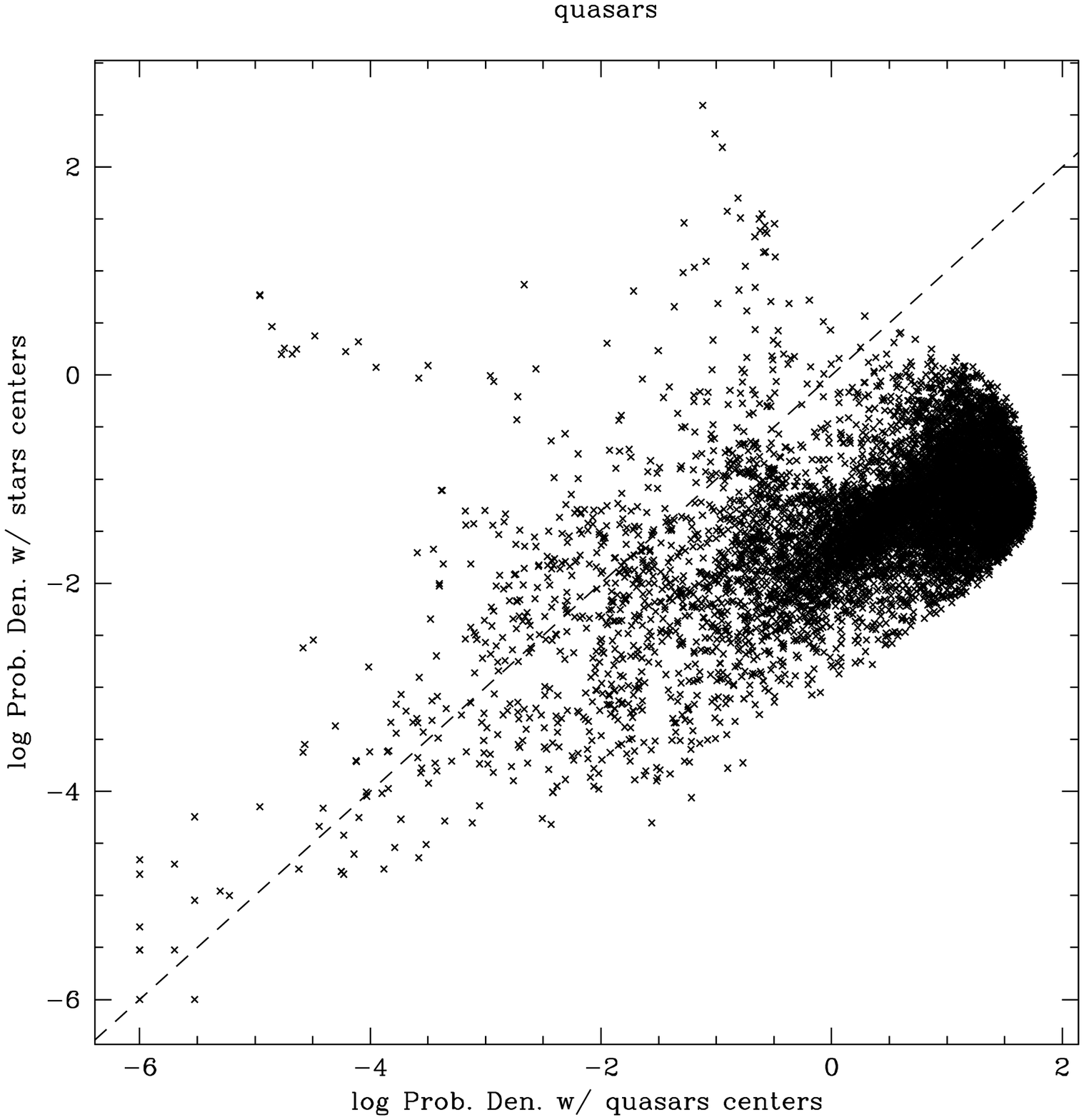}\\
\end{center}
{{\bf Figure 1:} Relative Likelihood of a SDSS source being a star or quasar
based on their observed colors. The 45 degree separation line is shown.
} 

\subsection{Quasar Target Selection}
We have also begun to use the Mixture Model algorithm to help in the selection
of quasars in multi-color space.  In Figure 1, we show a preliminary
implementation of such an algorithm using the SDSS data. Here, we have
clustered 8833 spectroscopically--confirmed SDSS quasars and 9999
SDSS stars (selected to be point--like objects) in 4D color--space
($u'-g'$,$g'-r'$,$r'-i'$,$i'-z'$) to obtain two pdf's; one for quasars and the
other for stars. Then given a new SDSS source with measured colors, one can
easily compute the relative likelihood that it is a star or quasar. As
illustrated in Fig. 1, we can achieve a high success rate with 96\%
of the quasars having a quasar probability density larger than stellar
probability densities and 99\% of the stars having a stellar probability
density higher than quasar probability densities {\it i.e.} the dashed lines
in these figures.

We plan several major improvements to this technique. This includes {\it i)}
the addition of other parameters like star--galaxy separation probability,
magnitudes, radio and X--ray fluxes {\it etc.}; {\it ii)} the use of synthetic
quasar and star SDSS colors to ensure we are not biasing ourselves since the
observed data clearly includes the survey selection function; {\it iii)}
increased testing using significantly more spectroscopic and photometric data
from the SDSS. 

In addition, these applications of the mixture model algorithm have
highlighted the need for improvements to the core technology, specifically the
need for the algorithm to incorporate observational errors on the data points
being clustered to obtain the pdf's. This is traditionally ignored in such
computer science orientated algorithms but is vital when analysing real
astronomical data. We also need to develop and improve the visualization of
the mixture model. At present, this is woefully inadequate and is beginning to
hinder our ability to quickly interpret the results of our mixture model.
These improvements to the algorithm will require new computer science and
statistical research.

\section{False Discovery Rate}
\label{FDR}

In a recent paper by Miller et al. (2001a), we introduced the False Discovery
Rate (FDR) to the astronomical community. This is a new statistical procedure
for performing multiple hypothesis tests on data and has three key advantages
over more traditional methods like a ``3--sigma'' threshold or the Bonferroni
method: {\it i)} It has a higher probability of correctly detecting real
deviations between the model and the data; {\it ii} it controls a
scientifically relevant quantity -- the average fraction of false discoveries
over the total number of discoveries; {\it iii)} it can be trivially adapted
to handle correlated data.

We have recently used FDR to solve two astronomical problems. The first is the
detection of the acoustic oscillations (``Baryon Wiggles'') in the power
spectrum of matter in the local universe (see Miller et al. 2001a,b,c for the
full details of this discovery). In Figure 2, we show our detection of the
``Baryon Wiggles'' along with a comparison of our work with the recently
released CMB Balloon data (MAXIMA \& BOOMERANG). The agreement between these
two measurements is impressive and it is re-assuring that our detection of the
``Baryon Wiggles'' is fully consistent with the CMB at a $z\sim1000$. In
summary, the FDR procedure is a less conservative procedure than the more
traditional multiple hypothesis testing methodologies (like ``2 sigma''
thresholding) commonly used in Astronomy. This has allowed us to detect the
``Baryon Wiggles'' in the local universe with much fewer data. This
illustrates the power of using new statistical tools in this era of high
precision cosmology as we attempt to extract the maximum amount of information
from these future surveys.

\begin{center}
\includegraphics[width=4.0in]{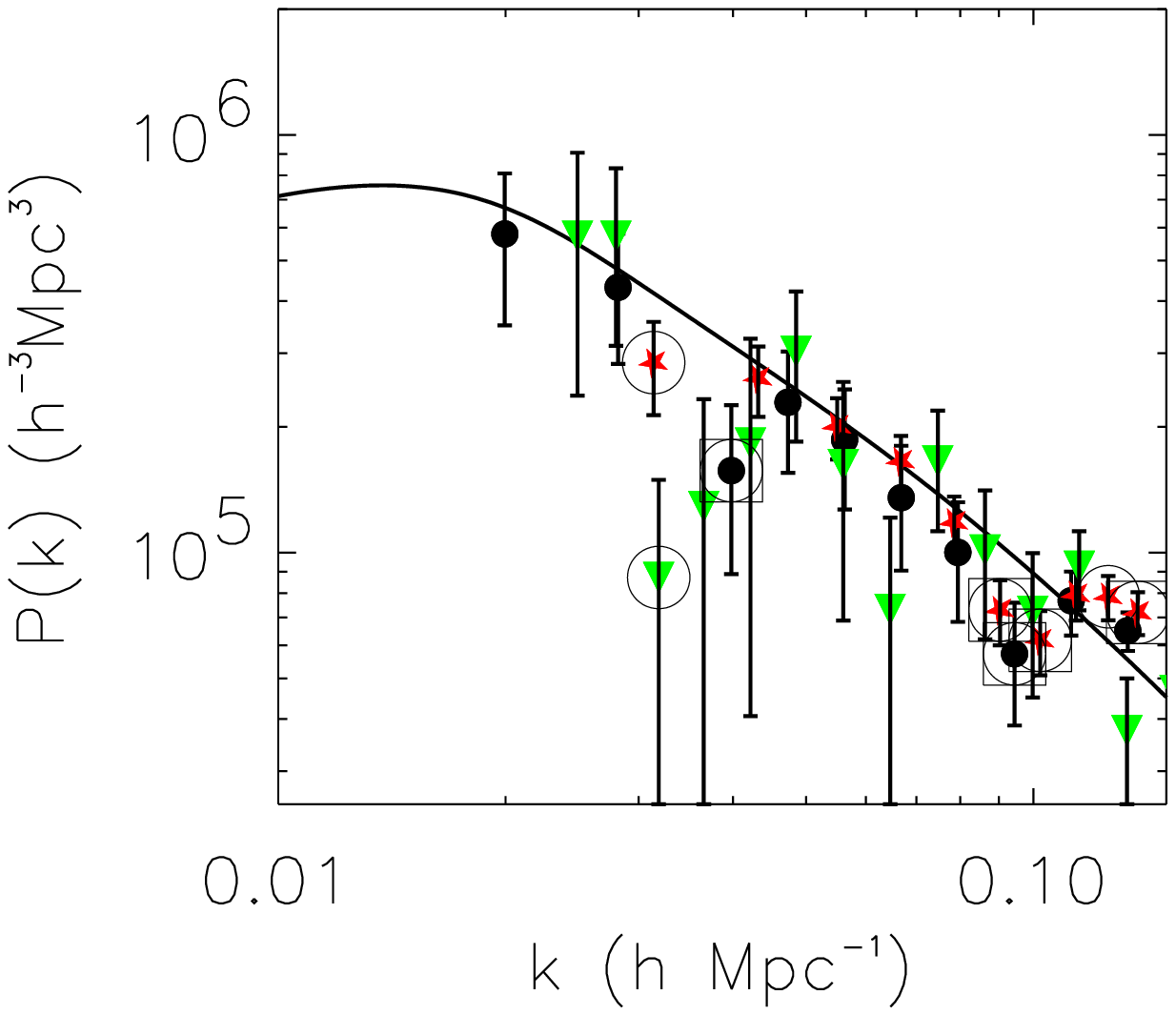}
\end{center}
{{\bf Figure 2:} Figure 5 taken from Miller et al. (2001a).  The figure shows
the amplitude--shifted power spectra for the three samples of uncorrelated
data (see Miller et al. 2001b for details). The points highlighted with a
circle denote rejections with $\alpha = 0.25$ (e.g. a quarter of the
rejections may be mistakes). The points highlighted by squares are for $\alpha
= 0.10$ (e.g.  a tenth of the rejections may be mistakes). The analysis
utilizes our best-fit model with the baryon wiggles removed as the null
hypothesis.  By controlling the false discovery rate, we can say with
statistical confidence that the two ``valleys'' are detected as features in
the power spectra.}\\

A second application of FDR is given in Hopkins et al. (2001) as part of a new
source detection algorithm for radio data. Specifically, Hopkins et al. (2001)
use FDR to determine which pixels in their radio telescope images are
consistent with sky noise or are part of a source. Traditionally, this is done
by apply a ``5 sigma'' threshold which, as discussed by Hopkins et al. (2001)
and Miller et al. (2001a), is a very conservative test. Hopkins et al. (2001)
compare the FDR method with Imsad and Sextractor (two traditional methods of
detecting sources in imaging data) and find it is significantly better than
these methods in detecting more, real sources without increasing the false
source detection rate.

\section{Bayes Network Anomaly Detector}
\label{Bayes}

Bayes Networks are a popular method for representing joint probability
distributions over many variables. The Bayes Nets have the advantage that
instead of using a single joint probability function (which can be prohibitive
since it may require a large number of parameters to fit the data), they
factor the distribution into a smaller number of conditional probability
functions for only a subset of the important variables.                                                   .
In practical terms, Bayes Nets have two limitations. They are computational
slow to learn and traditionally only work for discrete data. We have tackled
both of these issues using a new implementation of Bayes Networks called {\it
Mix-Nets} (see Davies \& Moore 2000) which uses the mixture model of Gaussians
to fit the data quickly over different subsets of the domain variables which
can then be combined into a coherent joint probability model for the entire
domain.  Once learned, the Bayes Net offers the ability to isolate sources with
a low likelihood of being produced by the model and this identifies those
sources as anomalies. Moreover, the Bayes Net provides the variables in the
joint probability model which cause this source to be anomalous.

We have used this technology to search for anomalies within the SDSS
photometric archive. Specifically, we have used 1.5 million SDSS detected
sources, each with 25 variables (magnitudes, sizes and shape parameters in all
5 of the SDSS passbands), to build a Bayes Network. We derive the overall
probability of each source (using the learned network) and rank the sources by
this probability. The bottom 1000 sources are flagged as anomalies and
visually inspected as they are unusual objects, within the data, based on the
joint probability model of these 25 attributes.

One of the major problems with this present approach is the existence of
errors within the data. At present, the most unusual objects are diffraction
spikes (around stars), asteroids and de--blending errors. This is
understandable since these errors have unphysical colors and shapes making
them gross outliers to many of the joint conditional probability
distributions. 

We plan to tackle this problem -- which is an issue of productivity -- using
an iterative loop where the scientist helps the Bayes Network focus on the
interesting astronomical anomalies. First, we will initially learn the Bayes
Network with all attributes and all data points of interest. The scientist
will be presented the bottom 1000 sources (the anomalies with the lowest
probability) and will interactively highlight obvious errors (like those
mentioned above).  As the Bayes Networks also stores the conditional
probabilities that caused this anomaly, we can use this information to
suppress further examples of such an error when we re--learn the Bayes Network
{\it i.e.} if diffraction spikes are always ``long'' and ``red'' we can use
that information to ignore further examples of this error. After a few
iterations, we should have interactively suppressed obvious errors based on
this feed--back loop and the scientist will be presented with a higher
percentage of physical anomalies. This is research in progress.

\begin{center}
{\includegraphics[width=2.3in]{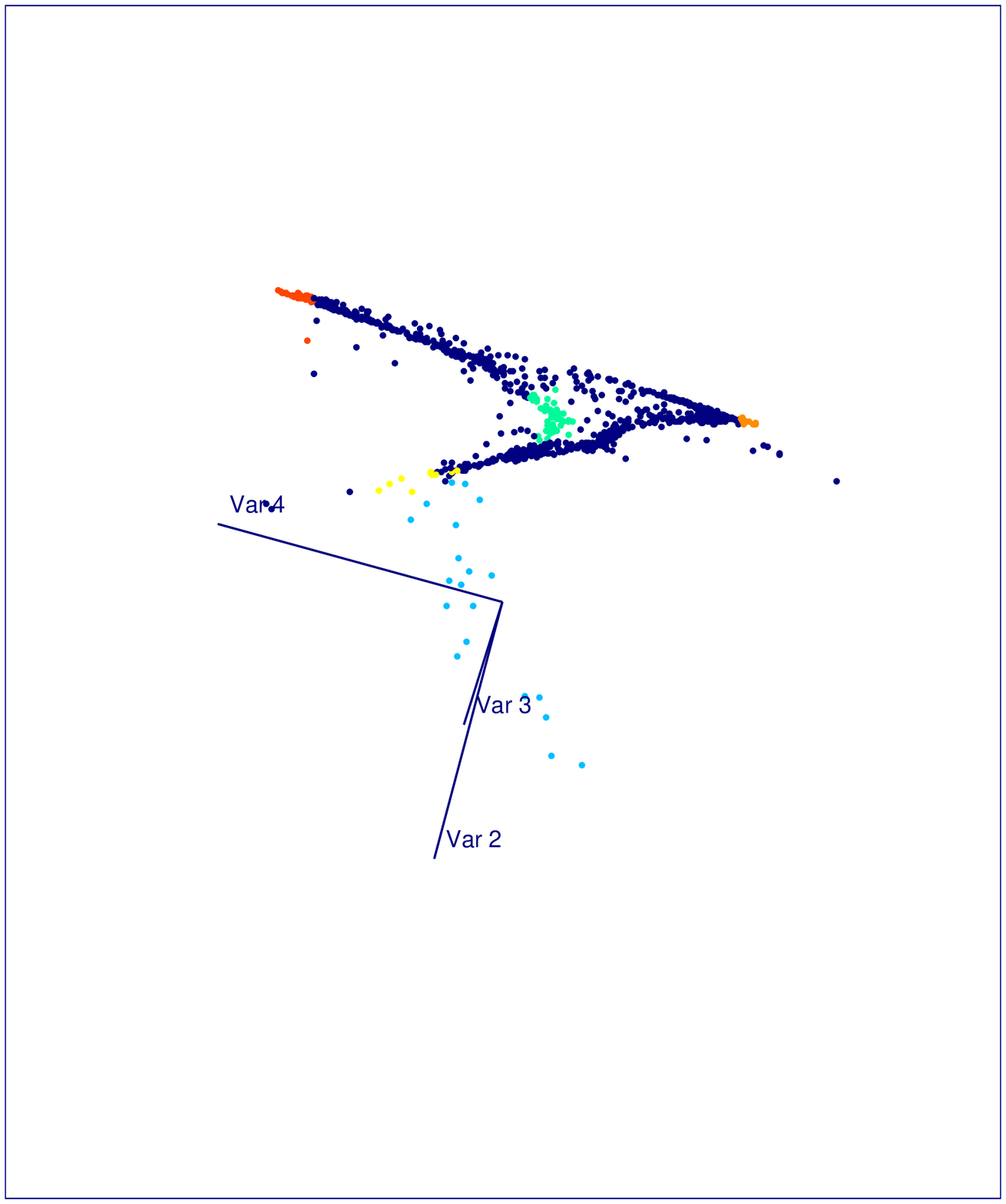}\includegraphics[width=2.3in]{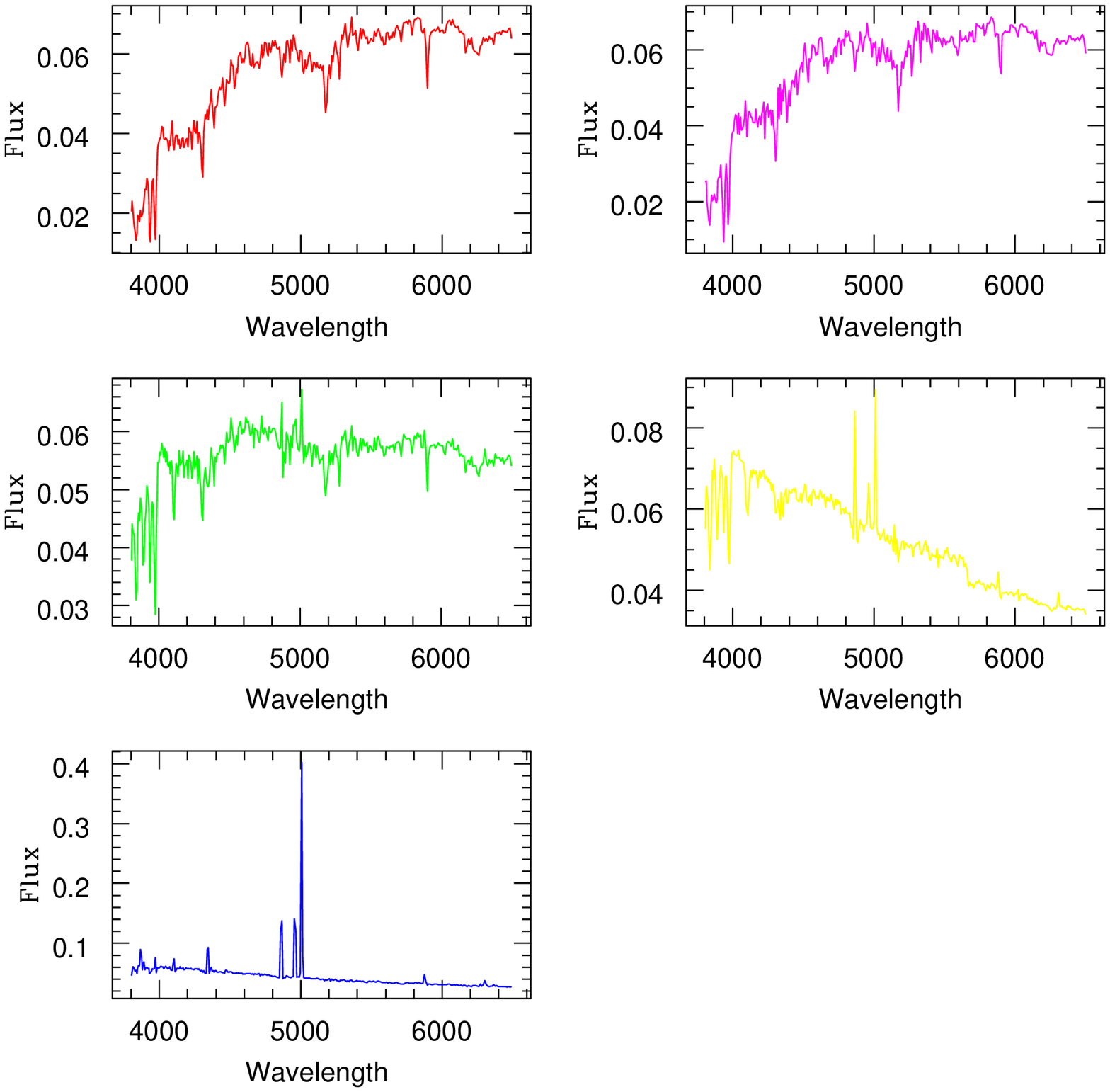}\\}
\end{center}
{{\bf Figure 3:} The LLE algorithm is applied to a sample of 500 galaxy
spectra (each of 2000 wavelength elements) in order to determine if galaxy
spectra occupy a lower dimensional subspace ({\it i.e.}\ if strong
correlations are present between the individual spectra). Using LLE to
compress this 500x2000 space down to a 3 dimensional subspace (see left panel
for the distribution of the coefficients for the 500 spectra in this 3D
space).  We find that the position of a galaxy within this subspace is
directly correlated with its spectral type (or, mean age of the galaxy). The
right panel shows the typical spectra associated with those points highlighted
in the left panel {\it i.e.} The red points go with the red spectrum {\it
etc.}. This simple example demonstrates how new computational techniques might
enable a radical compression in the dimensionality of physical data sets.  }

\section{Very High Dimensional Data}
\label{high}

The next generation of astronomical data will contain many thousands of
dimensions. This presents a new paradigm for data analysis techniques since
present algorithms and tools do not scale--up into such regimes. The handling
of very high dimensional data is an active research area in computer science
and statistics {\it e.g.} Isomap (Tenenbaum, de Silva \& Langford
2000) and LLE (Roweis \& Saul 2000). In Fig. 3, we show the power of
such algorithms through the use of LLE to non--parametrically study the
classification of SDSS spectra. 

\section{Conclusions}

In this paper, I have outlined an array of fast and efficient statistical
algorithms we are developing as part of the Pittsburgh Computational
AstroStatistics (PiCA) Group. This is a balanced, multi--disciplinary research
effort where all parties gain substantially from this cross--discipline
collaboration. For example, the fast algorithms enable new astrophysics to be
done and conceived ($N$--point functions), while the astrophysics problems
drive new computer science and statistics {\it e.g.} the incorporation of
errors into Bayes Networks and Mixture Models as well as new statistical
theory in extending FDR to slightly correlated data. Therefore, it is a rich
collaboration with many possibilities to simulate new and cutting--edge
research in computer science, statistics and astrophysics. This work is funded
in part through the NSF KDI and ITR programs and the NASA AISRP program and
makes use of SDSS data (see www.sdss.org). We acknowledge Don York for
carefully reading this manusript.

\section{References}
\noindent Connolly, A. J., et al. 2000, AJ (submitted), see astro-ph/0008187\\
Davies, S., Moore, A. W., 2000, {\it Proceedings of the Sixteenth Conference
on Uncertainty in Artificial Intelligence}\\
Hopkins A. M., et al. 2001, AJ (submitted)\\
Miller, C. J., et al. 2001a, AJ, see astro-ph/0107034\\
Miller, C. J., Nichol, R. C., Batuski, D.J., 2001b, ApJ, 555, 68\\
Miller, C. J., Nichol, R. C., Batuski, D.J., 2001c, Science, 292, 2302\\
Moore, A. W., 1991, Ph.D. Thesis, University of Cambridge\\
Moore, A. W., Lee, M. S., {\it Volume 8 of Journal of Artificial Intelligence Research}\\
Moore, A. W., et al., 2000, {\it Proceedings of MPA/MPE/ESO Conference "Mining the Sky}, see astro-ph/0012333\\
Nichol, R.C., 2000, {\it Proceedings from ``Virtual Observatories of the Future''} see astro-ph/0007404\\
Romer, A. K., 2000, ApJS, 126, 209\\
Roweis, S., Saul, L. K., 2000, Science, 290, 5500\\
Stocke, J. T., et al. 1991, ApJS, 76, 813\\
Stoughton, C., et al. in preparation\\
Szapudi, I., et al. 2001, ApJ, 548, 115\\
Tenenbaum, J. B., de Silva V., Langford, J. C., 2000, Science, 290, 5500\\
Turner, M. S., 2001, PASP, see astro-ph/0102057\\
Voges, W., et al. 1999, A\&A, 349, 389\\

\end{document}